\documentstyle[manuscript,aps]{revtex} 
\begin{document}
\draft
%\preprint{HEP/123-qed}

\title{ Collision Dynamics of Two Bose-Einstein 
Condensates in the Presence of Raman Coupling }

\author{Tao Hong, Tadao Shimizu}

\address{Department of Electronics and Computer Science, 
Science University of Tokyo in Yamaguchi, 1-1-1 
Daigaku-dori, Onoda, Yamaguchi 756-0884, Japan.}

%\date{August 8, 2000}
\maketitle

\begin{abstract}

A collision of two-component Bose-Einstein condensates 
in the presence of Raman coupling is proposed and studied 
by numerical simulations. Raman transitions are found 
to be able to reduce collision-produced irregular 
excitations by forming a time-averaged attractive optical
potential. Raman transitions also support a kind of 
dark soliton pairs in two-component Bose-Einstein 
condensates. Soliton pairs and their remnant single 
solitons are shown to be controllable by adjusting the 
initial relative phase between the two colliding 
condensates or the two-photon detuning of Raman 
transitions. 
\end{abstract}


%\narrowtext

\section{INTRODUCTION}

The realization of high-density Bose-Einstein 
condensates~\cite{BEC} invokes the study of nonlinear 
dynamics of matter waves and inventions of various 
optical techniques to manipulate Bose-Einstein 
condensates as well as constructions of various matter-wave interferometers for precision measurements. Usually, people would like to reduce irregular excitations and preserve the coherent property of high-density Bose-Einstein condensates as well as possible in many manipulating processes. However, due to the interactions between atoms, the strong repulsive interactions between Bose-Einstein condensates often lead to irregular excitations, especially in colliding processes. Inventing a convenient optical technique, which can reduce irregular excitations as well as exert effective controlling on the Bose-Einstein condensates, is very important.

Raman coupling is a very successful 
experimental technique in manipulating ultra-cold atomic 
gas between different internal and external states because it 
has the merit of avoiding spontaneous emission loss of 
atoms in transition processes. For example, recently 
several experiments have already successfully employed 
Raman coupling in manipulating Bose-Einstein 
condensates~\cite{Ramantransitions}. 

Here by proposing a Raman coupling scheme of atomic Bose-Einstein condensates, we show that Raman coupling can also be used to reduce irregular excitations produced in a colliding process of two Bose-Einstein condensates. Instead of irregular excitations, the Raman coupling produces a new type of coherent textures, namely dark soliton pairs in the colliding process. Additionally, we also find that these dark soliton pairs and their remnant dark solitons are very sensitive to the initial relative phase between Bose-Einstein condensates and the two-photon detuning of Raman transitions.
 
The paper is organized as follows. In Sec. II, we introduce the proposed colliding process of two Bose-Einstein condensates in the presence of Raman coupling, and the two coupled Gross-Pitaevskii equations for the description of this colliding process. In Sec. III, first we do some simplifications, then we show numerical simulations of the colliding process, analyze the function of Raman coupling in the colliding process, and discuss the physical origin and some special properties of dark soliton pairs. In Sec. IV, we draw conclusions as well as point out some potential applications of the proposed colliding process. 

\section{THEORETICAL MODEL}

First, we introduce the proposal in detail by using a schematic diagram in Fig.~\ref{fig1}. The black frame, in Fig.~\ref{fig1}(a), represents a transverse confining potential $V_{0}(x,y)$ for two atomic Bose-Einstein condensates 1 and 2 in the abstract. This potential can be formed, for example, by a blue-detuned hollow laser beam propagating along z-axis, which has already been demonstrated to be very effective in the recent experiment~\cite{Bongs}. Additionally, there are two red-detuned Gaussian laser beams 1 and 2 propagating along x-axis. The intensities of two laser beams are uniform in x and y directions, but in Gaussian shape in z direction, as indicated by the gray levels. Their frequencies are $\omega_{1}$ and $\omega_{2}$ respectively, as shown in Fig.~\ref{fig1}(b). As to the condensates, we assume that atoms have three internal states, $|g_{1}\rangle$, $|g_{2}\rangle$ and $|e\rangle$. Thus the two laser beams couple them together and form two Rabi transitions respectively, as shown in Fig.~\ref{fig1}(b). Because of the negative detuning ($\Delta<0$) and nonuniform intensities of the two laser beams, they can form two z-axial confining potentials for condensates 1 and 2 respectively, which are assumed to be initially distributed in states $|g_{1}\rangle$ and $|g_{2}\rangle$ respectively. Additionally, we assume that the laser beams 1 and 2 are initially separated by a distance and moving at velocities $v_{1}$ and $v_{2}$ respectively in opposite directions. Thus the two condensates are trapped by the two laser beams and moving at velocities same as those of laser beams in opposite directions. While the two laser beams overlap, the condensates in the overlap region are transferred between states $|g_{1}\rangle$ and $|g_{2}\rangle$ due to the occurrence of Raman transitions. As the two condensates go very close, they collide and interfere with each other. 

Next, we derive the equation for the description of this proposal. As shown in Fig.~\ref{fig1}(b), because atoms have three internal states, first we generally consider a condensate consisting of three components. To avoid spontaneous emission in Raman-transition process, we assume that the intermediate detuning $\Delta$ of laser beams 1 and 2 is much larger than other characteristic frequencies, such as the natural line width of the atomic transitions and the two-photon detuning $\delta$. As a result, the evolution of the $|e\rangle$-state component adiabatically follows the $|g_{1}\rangle$ and $|g_{2}\rangle$-state components, i.e.,
$\Psi_{e}=-({\Omega_{1}e^{-i\omega_{1}t}}\Psi_{g1}+
{\Omega_{2}e^{-i\omega_{2}t}\Psi_{g2}})/({2\Delta})$, 
where $\Psi_{e}$,$\Psi_{g1}$ and $\Psi_{g2}$ are the 
macroscopic wave functions of the three-component 
Bose-Einstein condensate in states 
$|e\rangle$, $|g_{1}\rangle$ and 
$|g_{2}\rangle$ respectively. Then, at zero 
temperature and under rotating-wave approximation, the 
light-coupled $|g_{1}\rangle$ and $|g_{2}\rangle$-state 
components can be described by the following coupled 
Gross-Pitaevskii equations: 
\begin{equation}
i\hbar\frac{\partial\Psi_{g1}}{\partial t}=
-\frac{\hbar^{2}}{2m}\nabla^{2} \Psi_{g1}
+V_{0}(x,y) \Psi_{g1}
+V_{g1}(z, t) \Psi_{g1}
+U_{0}|\Psi_{g1}|^{2}\Psi_{g1}
+U_{0}|\Psi_{g2}|^{2}\Psi_{g1}
+\hbar R(z, t)\Psi_{g2}
\label{GP1}
\end{equation}
\begin{eqnarray}
i\hbar\frac{\partial\Psi_{g2}}{\partial t}=
-\frac{\hbar^{2}}{2m}\nabla^{2}\Psi_{g2}
+\hbar \omega_{g}\Psi_{g2}
+V_{0}(x,y) \Psi_{g2}
+V_{g2}(z, t) \Psi_{g2}
+U_{0}|\Psi_{g2}|^{2}\Psi_{g2}   \\ \nonumber
+U_{0}|\Psi_{g1}|^{2}\Psi_{g2}
+\hbar R^{*}(z, t)
 e^{i(\omega_{2}-\omega_{1})t}\Psi_{g1}
\label{GP2}
\end{eqnarray}
where $m$ is the atomic mass. $U_{0}$ describes the atomic 
interaction strength, which is related to the s-wave 
scattering length $a_{sc}$ by $U_{0}={4 \pi \hbar^2 
a_{sc}}/{m}$. Here we assume that $a_{sc}>0$ and it is 
same for all collisions between atoms in same or different 
internal states. $V_{0}(x,y)$ denotes the transverse confining 
potential, which is common for both $|g_{1}\rangle$ and 
$|g_{2}\rangle$-state components. $V_{g1}(z,t)=\hbar 
|\Omega_{1}({\bf r},t)|^{2}/(4\Delta)$ and 
$V_{g2}(z,t)=\hbar |\Omega_{2}({\bf 
r},t)|^{2}/(4\Delta)$ are two localized optical potentials 
produced by the laser beams 1 and 2 
respectively. $R(z,t)= \Omega_{1}^{*}({\bf 
r},t)\Omega_{2}({\bf r},t)/(4\Delta)$ is the effective 
Rabi frequency of two-photon transitions. Because 
the laser beams 1 and 2 are in Gaussian shape and moving 
along z-axis, their corresponding Rabi frequencies can be 
written as $\Omega_{1}({\bf r},t) 
=\Omega_{10}e^{-(z-z_{01} -v_{1}t)^2/a_{1}^2}e^{-i{\bf 
k}_{1} \cdot {\bf r}}$ and $\Omega_{2}({\bf 
r},t)=\Omega_{20}e^{-(z-z_{02}-v_{2}t)^2/a_{2}^2} 
e^{-i{\bf k}_{2} \cdot {\bf r}}$, where $\Omega_{10}$ and 
$\Omega_{20}$ are maximum magnitudes of the two Rabi 
frequencies, $z_{01}$ and $z_{02}$ are initial center 
positions, $a_{1}$ and $a_{2}$ are half widths, and 
${\bf k}_{1}$ and ${\bf k}_{2}$ are wave vectors of the 
two laser beams respectively. Additionally, we assume that the wave 
vectors ${\bf k}_{1}$ and ${\bf k}_{2}$ are approximately 
equal, then the two-photon Rabi frequency can be written as
$R(z,t)= 
\Omega_{10}\Omega_{20}e^{-(z-z_{01}-v_{1}t)^2/a_{1}^2 
-(z-z_{02}-v_{2}t)^2/a_{2}^2} /(4\Delta) $. 
It is evident that when and only when the two 
potentials, $V_{g1}(z,t)$ and $V_{g2}(z,t)$, overlap with 
each other, the effective two-photon Rabi frequency becomes 
finite. All together, we reduce the description of 
the Bose-Einstein condensate from three components to two components now.

\section{NUMERICAL SIMULATION}

For simplicity and to give more prominence to the analysis of the affection of Raman transitions on the collision, we need to do some simplifications. We assume the transverse confining potential $V_{0}(x,y)$ is so tight and the z-axial confining potentials, $V_{g1}(z,t)$ and $V_{g2}(z,t)$, are so loose: that in the description of the transverse modes of the Bose-Einstein condensate, we can neglect the affection of $V_{g1}(z,t)$ and $V_{g2}(z,t)$ as well as interactions between atoms; that in the description of the longitudinal modes, we can find that the longitudinal dimension of the condensate is much larger than the healing length. We can also find that the transverse dimensions of the condensate are very small, and the time scale for adjustment of the transverse profile of the condensate to the equilibrium form appropriate for the instantaneous number of atoms per unit length is small compared with the time for an excited pulse to pass a given point. This is a low-density approximation, in fact, similar to that used in Ref.~\cite{Jackson}. Thus by letting $\Psi_{g1}({\bf r},t)=f_{g1}(z,t)g_{g1}(x,y)e^{-i\mu_{1}t/\hbar}$ and $\Psi_{g2}({\bf r},t)=f_{g2}(z,t)g_{g2}(x,y)e^{-i\mu_{2}t/\hbar}$ and through a deduction similar to that in Ref.\cite{Jackson}, we can get simplified one-dimensional Gross-Pitaevskii equations,

\begin{equation}
i\hbar\frac{\partial f_{g1}}{\partial t}=
-\frac{\hbar^{2}}{2m}\frac{\partial ^{2} f_{g1}}
{\partial z^2}
+V_{g1}(z, t) f_{g1}
+U_{0}^{'}|f_{g1}|^{2}f_{g1}
+U_{0}^{''}|f_{g2}|^{2}f_{g1}
+\hbar R(z, t) e^{-i(\mu_{2}-\mu_{1})t/\hbar}f_{g2}
\label{GP3}
\end{equation}
\begin{eqnarray}
i\hbar\frac{\partial f_{g2}}{\partial t}=
-\frac{\hbar^{2}}{2m}\frac{\partial ^{2} f_{g2}}
{\partial z^2}+\hbar \omega_{g} f_{g2}
+V_{g2}(z, t) f_{g2}    
+U_{0}^{'}|f_{g2}|^{2} f_{g2}
+U_{0}^{''}|f_{g1}|^{2} f_{g2}     \\ \nonumber
+\hbar R^{*}(z, t)
 e^{i(\omega_{2}-\omega_{1})t+i(\mu_{2}-\mu_{1})t/\hbar } f_{g1}
\label{GP4}
\end{eqnarray}
where $g_{g1}$ and $g_{g2}$ are the normalized transverse modes of the two components, $\mu_{1}$ and $\mu_{2}$ are the chemical potentials of $g_{g1}$ and $g_{g2}$ respectively, $f_{g1}$ and $f_{g2}$ are the longitudinal wave functions, $U_{0}^{'}=U_{0}\int |g_{g1}(x,y)|^{4}dxdy$ and $U_{0}^{''}=U_{0}\int |g_{g1}(x,y)g_{g2}(x,y)|^{2}dxdy$. 

As an initial condition for the colliding process, we assume that the two-component Bose-Einstein condensate is initially separated into two condensates, i.e.,  $\Psi_{g1}({\bf r},t)|_{t=0}$ and $\Psi_{g2}({\bf r},t)|_{t=0}$, which are trapped in the ground states of the two moving potentials, $V_{g1}(z,t)$ and $V_{g2}(z,t)$, respectively. For simplicity, we assume that the two potentials have same shapes, depths and speeds (but in opposite directions), thus the transverse and longitudinal modes of the two initial ground-state Bose-Einstein condensates are in same shape and the condensates has same numbers of atoms and same chemical potentials. 

Then we use split operator method to solve the above time dependent Gross-Pitaevskii Eqs.\ref{GP3} and \ref{GP4}. First, we show a typical colliding process in the presence of Raman coupling in Figs.~\ref{fig2}(a), (b) and Fig.~\ref{fig3}. For comparison, in Figs.~\ref{fig2}(c) 
and (d), we also show another different collision in 
which two condensates same as the above are trapped by 
two potentials similar to $V_{g1}(z,t)$ and 
$V_{g2}(z,t)$, but there are no Raman transitions between 
them. The elimination of Raman transitions is possible 
so long as the two laser beams couple the two ground states 
$|g_{1}\rangle$ and $|g_{2}\rangle$ with different 
excited states respectively instead of one common excited 
state. The parameter values used in the numerical 
simulation are as following: 
$\Omega_{10}^{2}/(4\Delta)=\Omega_{20}^{2}/(4\Delta)= 
\Omega_{10}\Omega_{20}/(4\Delta)=-6000/\tau$, 
$v_{1}=-v_{2}=9 a_{1}/\tau$, $a_{2}=a_{1}$, 
$U_{0}^{'}=U_{0}^{''}=2000 \hbar/\tau$, $z_{01}=-z_{02}=-a_{1}$, and 
the chemical potentials of initial ground-state 
Bose-Einstein condensates are both $\mu_{0} = 
-3.99\times 10^{3} \hbar/\tau$,  where $a_{1}$ and 
$\tau=2ma_{1}^{2}/\hbar$ are considered as the unit 
length and the unit time respectively.

In the collision without Raman transitions, the two 
condensates obstruct their passing through each other 
because of repulsive interactions between atoms, as shown 
in Figs.~\ref{fig2}(c) and (d). After the collision, we 
can see that small parts of condensates radiate out of 
the axial optical potentials, the remnant Bose-Einstein 
condensates oscillate violently in the potentials. The 
violent oscillations appear to be very irregular. In 
contrast, Figs.~\ref{fig2}(a), (b) and Fig.~\ref{fig3} 
show that while Raman transitions exist, the 
radiations become very weak, the violent oscillations of 
remnant Bose-Einstein condensates are much reduced. 
This indicates that the repulsive interaction between two Bose-Einstein condensates is greatly counteracted by an attractive optical potential produced by Raman transitions. We can understand the formation of this potential in this way. Because Raman transitions can transfer each condensate between two components, $f_{g1}$ and $f_{g2}$ circularly, each condensate interacts with the two axial optical potentials $V_{g1}$ and $V_{g2}$ alternatively. As a result, the light-shifted energy of each condensate, produced by the two laser beams, becomes a temporal combination of $V_{g1}$ and $V_{g2}$, and each condensate feels a vibrating optical potential locating at the collision center $z=0$. According to the overlap form of $V_{g1}$ and $V_{g2}$, the vibrating potential, in average, is attractive. Thus this time-averaged attractive potential counteracts the repulsive interaction between two condensates, so it becomes much easier for them to pass through each other. We therefore observe the oscillation amplitudes of condensates, produced by the repulsive interaction, become much small. But between $t=0.05\tau$ and $0.15\tau$, we can still observe that the volume of each condensate oscillates slightly in the colliding process due to the vibration of the attractive optical potential. 

In the colliding process, we also observe that dark 
solitons are generated, as shown in Fig.~\ref{fig3}. 
Formations of dark solitons are due to the interference 
of colliding condensates. Although the two condensates 
are initially distributed in two different internal 
states respectively, Raman transitions can transfer 
parts of them from one of the internal states to the other 
in the colliding process. Consequently, when the 
transferred part of one condensate overlaps with the part 
of the other condensate in the same internal state, 
interference fringes are produced. Due to nonlinear 
effect of condensates, interference fringes evolve into 
dark solitons. It is evident that this process is similar 
to that in Ref.~\cite{Scott}. 

However, solitons produced in this colliding process in 
the presence of Raman coupling are quite different from 
those single solitons described in 
Ref.~\cite{Jackson,Solitontheories,Solitonexperiments}. One of 
important differences is that these solitons always 
appear in pairs. As shown in Figs.~\ref{fig3}(a)-(d), in 
the colliding process, every soliton in one component of 
the condensates always has a corresponding soliton in the 
other component at the same spatial position. Another 
important difference is that the density difference 
between two components does not lead to a speed difference 
between two solitons in a pair. As shown in 
Figs.\ref{fig3}(a) and (b), sometimes the two components 
have quite different local densities at the location of 
a soliton pair, however, solitons in the pair always move 
synchronously in the condensates. The reason for synchronous motion of two solitons in a pair is that Raman transitions associate the two components tightly 
and consequently locks their phases together. As we can 
see in Fig.~\ref{fig4}, the phases of two components 
are almost completely same in the region of strong Raman 
transitions that are indicated by magnitudes of the normalized two-photon Rabi frequency $|4R\Delta/(\Omega_{10}\Omega_{20})|$. Dark solitons are phase kinks in essence, so 
there is no doubt that solitons, with similar spatial phases, in a pair move 
synchronously. Additionally, the cross phase modulation 
between dark solitons in a pair also makes a repulsive 
effect between them, and tends to separate them. It is 
evident that this repulsive effect is also counteracted 
by Raman transitions, which produce an equivalently 
attractive effect between solitons in the pair by locking 
their phases together. Even after a collision of two 
soliton pairs A and B, solitons in each pair can still 
preserve their waveforms and trajectories well, as shown 
in Figs.~\ref{fig3}(b)-(c) and Figs.~\ref{fig2}(a)-(b) 
respectively. This illustrates that collisions 
between soliton pairs are elastic collisions, which are 
usually considered as an important common characteristic of all kinds 
of solitons. The synchronous motion of solitons in the pair also indicates the propagation speed of the soliton pair is determined by the total local density of the two components instead of each one component. We can find this property from Figs.~\ref{fig2}(a)-(b). The propagation speed of the soliton pair, in the region $z<0$, at time $t=0.078\tau$ is larger than that at time $t=0.047\tau$, because the total density of the two components is increased in the overlap region,. Some similarities between a soliton pair and a single soliton can also be seen. For example, the propagation speed, the width and the density contrast of a soliton pair also depend on the value of its phase kink that is also varied by the local density gradient. We can find these properties by a careful observation of Figs.~\ref{fig2}(a)-(b), Fig.~\ref{fig3} and Fig.~\ref{fig4}. Because the density of the condensates varies in parabola-like form in space, we can find its phase kink flips in the boundary region of the condensates, simultaneously its speed flips too, and as a result, the soliton pair oscillates inside the condensates.

The two components of the condensates begin to separate 
very apparently after $t=0.15\tau$, because of the 
separation of the two axial optical potentials, as shown 
in Figs.~\ref{fig2}(a)-(b). Simultaneously, Raman 
transitions become weaker and weaker because of the 
decrement of the overlap of the two laser beams, as shown 
in Figs.~\ref{fig3}(d)-(f). As a result, soliton pairs 
become unstable and are to be separated in this process. 
Critically depending on the local evolution of the two components, 
some solitons disappear and some of them remain 
and propagate into the separated one-component 
condensates, as shown in Figs.~\ref{fig2}(a)-(b) and 
Figs.~\ref{fig3}(e)-(f). Because of the disappearance of 
cross phase modulation in each one-component condensate, the 
oscillation behaviors of remnant single solitons are 
different from those of soliton pairs. Especially, the 
oscillation periods of remnant single solitons appear 
to be quite different from those of previous soliton pairs.

The generation of dark soliton pairs is controllable. 
Usually, the relative phase between two condensates 
in different internal states is 
meaningless because of the orthogonality of the internal 
states. However, here Raman transitions can form a circulation 
of atoms between the two internal states, the relative phase between the two condensates are related and therefore becomes very 
important. Because the formation of dark soliton pairs 
is due to the interference of colliding condensates, we 
can also control the generation of soliton pairs as well as 
remnant single solitons by adjusting the initial 
relative phase between two colliding condensates or the 
two-photon detuning of Raman transitions, as shown in 
Fig.~\ref{fig5}.

In the above, we have discussed the collision of two Bose-Einstein condensates under the one-dimensional approximation, then will a three dimensional collision be completely different from the one-dimensional collision? Some difference must exist. For example, in three-dimensional process, dark soliton pairs might be distorted by transverse perturbations and evolve into vortices because similar behaviors of single dark solitons have been found \cite{Solitonexperiments}. However, in the above analysis, we have already found that the function of Raman coupling in the colliding process is, in fact, to lock the phases of the two components of the condensates by forming the strong atom circulation between the two internal states, and as a result, solitons appear in pairs. As for vortices, they are phase singular points, so they are quit similar to dark solitons in essence. We think that Raman coupling can also lock the phases of two components of the condensates in a three-dimensional case, and vortex pairs might therefore be formed. Certainly, more affirmative answer should be gotten in a numerical simulation of a three-dimensional collision, which will be done in the future work.

\section{CONCLUSION} 

In conclusion, we have proposed a new Raman coupling 
scheme of Bose-Einstein condensates, and analyzed the 
influence of Raman transitions on the collision and 
excitation of two-component Bose-Einstein condensates. 
We have found that Raman transitions can reduce 
collision-produced irregular excitations by forming a 
time-averaged attractive potential. Raman transitions 
also support a new kind of dark solitons, i.e., dark 
soliton pairs in two-component Bose-Einstein condensates, 
by locking the phases of the two components. We have also 
shown the control of soliton pairs and their remnant 
single solitons by adjusting the initial relative phase 
between the two colliding condensates and the two-photon 
detuning of Raman transitions. The dynamical sensitivity of the dark soliton pairs and remnant solitons to these parameters indicates that the proposed collision can be used in the measurement of relative phase between different high-density Bose-Einstein condensates. Currently, people are considering constructions of various interferometers with matter waves for high precision measurements. We think that this proposal might provide a useful method to this area.

\acknowledgments
T. Hong thanks Japan Science Promotion Society for the 
partial financial support. This work is supported by a 
Grant-in-Aid for JSPS Fellows from the Ministry of 
Education, Science, Sports and Culture of Japan.

\begin{figure}
\caption{(a) Collision and Raman coupling scheme of two atomic Bose-Einstein condensates. (b) Atomic internal energy levels of 
Bose-Einstein condensates and light coupling. 
$|g_{1}\rangle$, $|g_{2}\rangle$ denote two ground 
states and $|e \rangle$ denotes an excited state, energy 
eigen values of which are $0$, $\hbar\omega_{g}$ and 
$\hbar\omega_{e}$ respectively. The laser beams 1 and 2, with 
frequencies $\omega_{1}$ and $\omega_{2}$ respectively, couple these 
states together, forming so-called ``$\Lambda$'' 
configuration Raman transitions. $\Omega_{1}$ and 
$\Omega_{2}$ are corresponding single-photon Rabi 
frequencies. The intermediate detuning and two-photon 
detuning of Raman transitions are 
$\Delta=\omega_{1}-\omega_{e}$ and 
$\delta=\omega_{1}-\omega_{2}-\omega_{g}$ respectively. 
Direct transitions between $|g_{1} \rangle$ and 
$|g_{2} \rangle$ are electric-dipole forbidden.}
\label{fig1}
\end{figure}

\begin{figure}
\caption{Contour plots of time evolution of the 
condensate wave functions $|f_{g1}|$ and 
$|f_{g2}|$ in two colliding processes. (a) is for 
$|f_{g1}|$ and (b) is for $|f_{g2}|$ in the 
collision with Raman coupling. (c) is for 
$|f_{g1}|$ and (d) is for $|f_{g2}|$ in the 
collision without Raman coupling. In these collisions, 
the initial phases of $f_{g1}$ and $f_{g2}$ are 
$\pi/4$ and $0$ respectively, and the detuning $\delta=0$. 
The dash-dot lines denote the boundaries of the 
corresponding axial optical potential, $V_{g1}(z,t)$ or 
$V_{g2}(z,t)$. These boundaries correspond to the 
boundaries of Gaussian laser beams at the $1/e$ of maximum 
intensities. Between $t=0.05\tau$ and $0.15\tau$, there are four
zigzag density dip traces in (a) and (b). 
These four density dip traces indicate the oscillations 
of four dark solitons inside the condensates.}
\label{fig2}
\end{figure}

\begin{figure}
\caption{Generation of dark soliton pairs in the 
collision with Raman coupling and remnant single 
solitons. The thin solid lines denote $|f_{g1}/f_{0}|$, 
the thick solid lines denote $|f_{g2}/f_{0}|$, and 
the dashed lines denote 
$|4R\Delta/(\Omega_{10}\Omega_{20})|$. $f_{0}$ is 
the maximum of $|f_{g1}|_{t=0}$ and 
$|f_{g2}|_{t=0}$. There are four dark solitons in each figure 
of (a), (b) and (c). The arrow A marks one soliton pair, 
and the arrow B 
marks another soliton pair respectively. The 
values of $t$ denote the time of subfigures sampled from 
the colliding process shown in Fig.\ref{fig2}(a) and 
(b).}
\label{fig3}
\end{figure}

\begin{figure}
\caption{Phase evolution of the condensate wave functions in the collision with Raman coupling. The solid lines in (a)-(c) denote phase of $f_{g1}$, the solid lines in (d)-(f) denote phase of $f_{g2}/f_{g1}$, and the dashed lines denote 
$|4R\Delta/(\Omega_{10}\Omega_{20})|$. The arrow A marks the position 
of one soliton pair, and the arrow B marks the position of the other
 dark soliton pair respectively. The 
values of $t$ denote the time of subfigures sampled from 
the colliding process shown in Fig.\ref{fig2}(a) and 
(b).}
\label{fig4}
\end{figure}

\begin{figure}
\caption{Contour plots of time evolution of the 
condensate wave functions $|f_{g1}|$ and 
$|f_{g2}|$ in two colliding processes. (a) is for 
$|f_{g1}|$ and (b) is for $|f_{g2}|$ in one 
collision. In this collision, the initial phases of 
$f_{g1}$ and $f_{g2}$ are $\pi/2$ and 
$0$ respectively, and the detuning $\delta=0$.  (c) is 
for $|f_{g1}|$ and (d) is for $|f_{g2}|$ in the 
other collision. In this collision, the initial phases 
of both $f_{g1}$ and $f_{g2}$ are $0$s, and the 
detuning $\delta=100/\tau$. The dash-dot lines denote 
the boundaries of the corresponding axial optical 
potential, $V_{g1}(z,t)$ or $V_{g2}(z,t)$. These 
boundaries correspond to the boundaries of Gaussian laser 
beams at the $1/e$ of maximum intensities.}
\label{fig5}
\end{figure}

\end{document}